\documentclass[a4paper]{jpconf}
\usepackage{graphicx}
\usepackage{amsmath}
\usepackage{amsbsy}
\usepackage{amsfonts}
\usepackage{amssymb}
\usepackage{bm}
\usepackage{url}

\usepackage{placeins}
\usepackage{xcolor}
\definecolor{orangered}{HTML}{FF7110}
\definecolor{green}{HTML}{00CC00}

\begin{document}
\title{The structure of magnetic turbulence in the heliosheath region observed by \textit{Voyager 2} at 106 AU}

\author{Federico Fraternale}
\address{Dipartimento di Scienza Applicata e Tecnologia, Politecnico di Torino, Torino, 10131,  Italy}
\author{Nikolai V Pogorelov}
\address{Department of Space Science, University of Alabama in Huntsville, Huntsville, AL 35805, USA}
\author{John D. Richardson }
\address{Kavli Institute for Astrophysics and Space Research, Massachusetts Institute of Technology, Cambridge MA, 02139, USA}
\author{Daniela Tordella}
\address{Dipartimento di Scienza Applicata e Tecnologia, Politecnico di Torino, Torino, 10131,  Italy}

\ead{federico.fraternale@polito.it}

\begin{abstract}
It is currently believed that the turbulent fluctuations pervade the outermost heliosphere. Turbulence, magnetic reconnection, and their reciprocal link may be responsible for magnetic energy conversion in these regions. The governing mechanisms of such anisotropic and compressible magnetic turbulence in the inner heliosheath (IHS) and in the local interstellar medium (LISM)  still lack a thorough description. 
The present literature mainly concerns large scales which are not representative of the inertial-cascade dynamics of turbulence. Moreover, lack of broadband spectral analysis makes the IHS dynamics remain critically understudied. Our recent study \cite{fraternale2019a} shows that 48 s magnetic-field data from the \textit{Voyager} mission are appropriate for a spectral analysis over a frequency range of six decades, from $5\times10^{-8}$ Hz to $10^{-2}$ Hz. Here, focusing on the \textit{Voyager 2} observation interval from 2013.824 to 2016.0, we describe the structure of turbulence in a sector zone of the IHS. A spectral break at about $7\times10^{-7}$ Hz (magnetic structures with size $\ell\approx1.3$ Astronomical Units) separates the energy-injection regime from the inertial-cascade regime of turbulence.  A second scale around $6\times10^{-5}$ Hz ($\ell \approx0.017$ AU) corresponds to a peak of compressibility and intermittency of fluctuations.


\end{abstract}

\section{Introduction}

The \textit{Voyagers} (V1, V2) are the only operating spacecraft providing us with \textit{in situ} data from the outermost part of heliosphere. The inner heliosheath (IHS) is the region of space between the termination shock (TS) and the heliopause (HP). The HP is a tangential discontinuity that separates the solar wind (SW) from the local interstellar medium (LISM).

Major scientific questions are related to the physical processes responsible for the conversion of magnetic energy and SW heating, acceleration and transport of energetic particles \cite{mccomas2006,fisk2009,heerikhuisen2010,pogorelov2016,zhao2017}, existence and topology of the sector (swept by the heliospheric current sheet, HCS) and unipolar regions of the IHS \cite{hill2014,richardson2015}, and coupling between the interstellar and heliospheric magnetic fields \cite{pogorelov2017b}. For details, we refer the reader to three comprehensive reviews of the heliosheath plasma and related physical processes \cite{richardson2013,zank2015,pogorelov2017a}. 

These topics are tightly linked to the turbulent nature of the IHS/LISM plasma and magnetic fields \cite{burlaga2006,burlaga2006b,burlaga2009,burlaga2015,burlaga2018,fraternale2019a}, and the potential presence of magnetic reconnection \cite{drake2010,hill2014,drake2017,pogorelov2013b,pogorelov2017b}.  Heliospheric numerical simulations are shedding light on the three-dimensional heliospheric structure  (see, e.g. \cite{pogorelov2012,pogorelov2013a,pogorelov2015,pogorelov2017b,kim2017}). 
Notably, taking into account the solar-cycle variations made it possible to reproduce many observed features of the IHS and LISM bulk plasma flow and magnetic field. Moreover, it has been shown that the transition to chaos is possible, and the turbulence may be responsible for the SW heating and the observed average values of the heliospheric magnetic field \cite{pogorelov2013a,pogorelov2013b}. However, resolving the inertial range of turbulent fluctuations (say, $\ell\lesssim 1.5$ AU)  numerically is still unfeasible except for a very small computational domain, which makes the spectral analysis of \textit{Voyagers}' data over a broad range of spatial and temporal scales highly desirable to improve and constrain the models. Unfortunately, the intrinsically space- and time-local,  one-dimensional nature of spacecraft measurements, lack of plasma data at V1, presence of data gaps, and high level of noise in general, make such analysis nontrivial.

This study extends our recent work \cite{fraternale2019a}, where a spectral analysis of V1 and V2 data magnetic field data was performed for several IHS and LISM periods (in both unipolar and sector zones). It was made for a range of scales unprecedented in the literature (spacecraft-frame frequencies $10^{-8}<f<10^{-2}$ Hz, corresponding spatial scales  $5\times10^{-5}<\ell<65$ AU). Following the same line of research, here we focus on the most recent V2 data publicly available. In particular, we analyze the V2 magnetic field measurements from   2013.824 (day-of-year 300) to 2016.0.  The plasma and magnetic fields observed by V2 during 2015 have been discussed in details in \cite{burlaga2018b}. Previously, in \cite{burlaga2017}, it was shown that near 2013.824 at 103 AU V2 entered the IHS sector region, which was likely due to the increasing latitudinal extent of the HCS related to the near-maximum solar conditions.  Here, we investigate the spectral properties of magnetic field fluctuations in the \textit{energy-injection} and  \textit{inertial-cascade} ranges of turbulence, with focus on the variance anisotropy, the presence of compressible modes, and high-order multi-scale statistics. 

In section \ref{sec:data} we present details of the data set used for this study and the methodology adopted for multi-scale analyses. In section \ref{sec:results} results are shown: \ref{sec:energy_inj_regime} discusses the energy-injection regime of fluctuations and \ref{sec:interialcascade_regime} is focused on the inertial-cascade of turbulence. 

We believe that these analyses will be of importance for the improvement of existing theoretical and numerical IHS models.

\section{Data}\label{sec:data}
\begin{figure}
\centering
\includegraphics[width=0.75\textwidth]{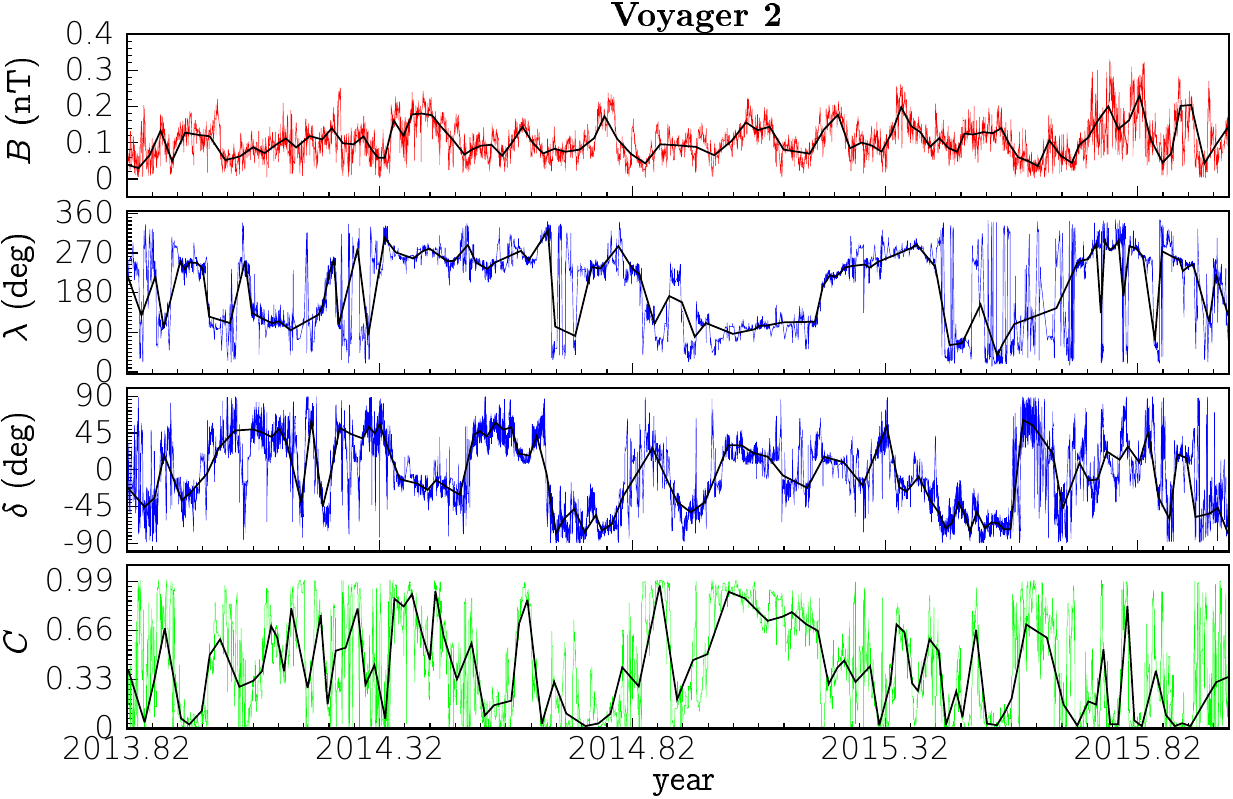}
    \caption{\textit{Voyager 2} data considered in the present study, observation period 2013.824--2016.0. Top to bottom: the magnetic-field magnitude $B=|\mathbf{B}|$;  the azimuthal angle $\lambda=\tan^{-1}({B_\mathrm{T}}/ {B_\mathrm{R}})$; the elevation angle $\delta=\sin^{-1}({B_\mathrm{N}}/ {B})$; and the fraction of fluctuating magnetic energy in the B-parallel direction, computed as $C=\left\langle\left[{\mathbf{ B_0}\boldsymbol{\cdot}\boldsymbol{\delta}\mathbf{B}}{/ (B_0  \delta B)}\right]^2\right\rangle$ ($\delta \mathbf{B}=\mathbf{B}-\mathbf{B}_0$). Data points with $|B_j|<0.03$ nT  have been disregarded in the computation of $\lambda$ and $\delta$. (High-resolution data is taken from the COHO website \url{https://cohoweb.gsfc.nasa.gov/coho/}). \label{fig:data}  }
\end{figure}

In the time interval from 2013.824 to 2016.0, V2 was traveling at an heliocentric distance of 106.5$\pm$3.4 AU, latitude of $-30.5^\circ$ and longitude $-217.5^\circ$ in HelioGraphic Inertial (HGI) coordinates.  Since the sector-boundary crossing in 2013.824, V2 has been traveling inside the sector region of the IHS \cite{richardson2016b}. This study considers magnetic field \textit{in situ} data provided by the V2 MAG experiment \cite{behannon1977}. The field's magnitude and angles are shown in Figure \ref{fig:data}. To explore a broad range of scales, we used data at the highest sampling rate publicly available, i.e., the 48 s resolution (processed data can be downloaded from the NASA's Space Physics Data Facility  \url{https://spdf.gsfc.nasa.gov/}). Data are provided in the HGI RTN coordinate reference system, but for the purpose of this study it was convenient to adopt a reference frame having one axis aligned with the average field $\mathbf{B_0}$ (the $\parallel$ axis). 
The $\perp_1$ and the $\perp_2$ axes form the plane orthogonal to $\mathbf{B_0}$, with $B_{\perp 1}$ belonging  the T-N plane. Given that in the IHS $\mathbf{B_0}$ is nearly aligned with the T direction, it follows that $\perp_1$ is approximately aligned with N and $\perp_2$ with R.

Computing the power spectral density (PSD) is challenging due to the sparsity of the \textit{Voyager} data set. In the interval considered in this study,  72\% of 48 s data points are missing. Typically, the data gaps of 8--16 hours in the IHS are largely due to ground tracking issues and occur once a day. 
The average frequency of the largest gap is $f_\mathrm{gap}=2.4\times10^{-5}$ Hz.
The PSD is computed on the basis a comparative analysis of four spectral estimation methods (Compressed Sensing CS, linear interpolation CI, optimization of model spectra OP, Fourier transform of gap-free subsets SS). 
These techniques have  been successfully adopted already in our previous papers on the SW turbulence at 5 AU \cite{fraternale2016,iovieno2016,gallana2016}, and in the recent study  \cite{fraternale2019a} regarding IHS and LISM turbulence. CS and SS has also been used in \cite{sorriso2019} to analyze turbulence in the Earth's magnetosphere, in proximity of the magnetopause. Scripts and detailed descriptions of all methods are provided in the Supplemental Material of \cite{gallana2016} and in Chap. 4 of  \cite{fraternale2017phd}. Some tests specific to IHS data can be found in Appendix A of \cite{fraternale2019a}.  

The accuracy of \textit{Voyager} observations is affected by the level of noise. This includes the magnetometer's sensibility (0.006 nT), and various sources of noise such as the telemetry system, the interference with other instruments, and the calibration process. The resulting 1$\sigma$ error of magnetic field components is estimated to set around 0.03 nT \cite{berdichevsky2009}. The actual level and distribution of the noise are unknown and may differ from the above estimate. If a white-noise model with the amplitude of 0.03 nT is considered, the PSD is constant and equal to 0.029 nT$^2$s. This threshold is represented in all figures with a gray band. It can be noticed, however, that the actual level of noise may be lower for the period considered here, since a serious spectral flattening is not observed until $f\approx10^{-3}$ Hz. In fact, this frequency intercepts the spectra at $P\approx10^{-3}$ nT$^2$s, which would correspond to a white noise of 0.0056 nT amplitude (see Figure \ref{fig:PSD}(a)). Consequently, as discussed in \S \ref{sec:results}, the present spectral analysis may be affected by the noise in the frequency range $4\times10^{-4}\lesssim f\lesssim 10^{-2}$ Hz, which likely includes the transitional MHD-to-kinetic regime of magnetic turbulence. 

\textcolor{black}{Higher-order statistics of magnetic field increments (the structure functions, in the time domain) are computed from both 48 s data and 1824 s averages, which helps to estimate the effect of noise. The p-th structure function of the j-th magnetic field component is defined as $S_{p,j}(\tau)=\left\langle |\Delta B_j|^p \right\rangle$, where $\Delta B_j(t;\tau)=B_j(t)-B_j(t+\tau)$ is the magnetic field increment of the j-th component for the time lag $\tau$, and angle brackets denote the time-average for the period analyzed (as usual, this assumes ergodicity of the underlying physical processes). We computed $S_{p,j}$ from available data points as }


\begin{gather}\label{eq:strucfun2}
S_{p,j}(\tau_k)=\frac{1}{N(\tau_k)}\sum\limits_{i=1}^{N(\tau_k)} |B_j(t_i)-B_j(t_i+\tau_k)|^p,\\ \tau_k=k\cdot\Delta t_s\ \   k=1,\dots,n, \nonumber
 \end{gather}
\textcolor{black}{where $\Delta t_s=48$ s (or 1824 s), and $n=401600$ (or 10200) is the total number of points of the data set. $N(\tau)$ is the number of the available increments $\Delta B(t_i,\tau)$. Note that for contiguous data sets $N(\tau)$ is necessarily a linearly-decreasing function of the time lag, due to the limitedness of the data sequence. In the case of \textit{Voyager} data sets in the IHS, however, due to data gaps $N(\tau)$ is also oscillating, with minima at multiples of $\tau_\mathrm{gap}=1/f_\mathrm{gap}\approx11.5$ hours, that corresponds to the periodicity of large gaps.} Points with smaller statistical significance (when $N(\tau)< 0.25 \max[N(\tau')],\ \tau'\in[\tau-48h,\ \tau+48h]$)) have been disregarded in the computation of power-law exponents (Table \ref{tab:fluc_stats}) and are shown in gray in Figure \ref{fig:SF}(a). \textcolor{black}{The figure will also reports the fluctuation intensity ratio $\mathcal{I}(\tau)=\langle |\boldsymbol{\Delta} \mathbf{B}|/B_\tau \rangle$ between the  the magnitude of the increments defined above and the average magnitude between the instant t and $t+\tau$, $B_\tau(t;\tau)$.} 

\section{Results and discussion}\label{sec:results}

Figure \ref{fig:PSD}(a) shows the power spectral density (PSD, or $P$) of the magnetic field components. The red, green and blue curves stand for $\delta B_\parallel,~\delta B_{\perp1},~\delta B_{\perp2}$, respectively, while the total magnetic energy $E_m(f)=P[B_\parallel]+P[B_{\perp1}]+P[B_{\perp2}]$ is represented in black. Figure \ref{fig:PSD}(b) shows the spectral variance anisotropy and a proxy for spectral compressibility. The former is defined as $A_j(f)=P[B_j]/E_m$, while the latter is $C(f)=P[|\mathbf{B}|]/E_m$, the ratio between the spectrum of the magnetic field magnitude and the trace.  It is considered here as a proxy for the density fluctuations, as they are typically strongly correlated with the fluctuations of magnetic field magnitude \cite{roberts1987a}. Note that most of notations and definitions that we use in the present work are the same as in \cite{fraternale2019a}.  

Inevitably, single-spacecraft measurements cannot provide the omni-directional spectrum but only the reduced one (1D), containing the contribution of all vector wavenumbers. Since  the spacecraft speed during the analyzed period is about 0.1 of the bulk wind speed ($V_\mathrm{SW}\approx150$ km/s) and about 0.3 of the Alfv{\'e}n speed ($V_\mathrm{A}\approx53$ km/s), we used the Taylor's hypothesis to convert the spacecraft-frame (SC) frequencies to wavenumbers \cite{taylor1938}. The average magnetic field being nearly orthogonal to the wind direction, such wavenumbers can be interpreted as perpendicular to $\mathbf{B_0}$, $k_\perp\approx2\pi f_\mathrm{SC}/V_\mathrm{SW}$. This information can be used to compare the present results with theoretical findings on the spectral scaling laws of anisotropic turbulence \cite{zhou2004}, and to estimate the order of magnitude of magnetic structures in the direction of the wind  (the average azimuthal and elevation angles of the thermal plasma flow are $\lambda_v=43^\circ$ and $\delta_v=-25^\circ$, respectively). However, it should be reminded that the application of the Taylor's hypothesis in the IHS is more critical than in the supersonic SW upstream the termination shock, and that it may not be applicable in the high-frequency range of the spectrum, especially if the dispersive waves play an important role.

Table \ref{tab:averages} offers the average quantities, typical length scales, and frequencies. Table \ref{tab:fluc_stats} summarizes the results of the fluctuation analysis. The left panel of Figure \ref{fig:SF} shows the structure functions of magnetic field increments for parallel fluctuations ($S_{p,\parallel}$, red curves) and perpendicular fluctuations ($S_{p,\perp}=[S_{p,\perp1}+S_{p,\perp2}]/2$, blue curves). \textcolor{black}{In the insert of panel (a), the fluctuation intensity ratio $\mathcal{I}$ is shown}. The right panel shows the scale-dependent kurtosis of magnetic increments for the three field components, $K_j=S_{4,j}/S_{2,j}^2$,  a measure of intermittency. 

\begin{table*}
\lineup
\caption{Average quantities, in the V2 observation period 2013.824--2016.0. \label{tab:averages}}
\begin{indented}
\centering
		\item[]\begin{tabular}{@{}llll}
			\br
		\centre{2}{Parameter} & \centre{2}{Value}                      \\
			\mr
			\boldmath $d_\mathrm{SC}$        & Spacecraft-Sun radial  distance                        & 106.5$\pm$3.44  & AU         \\

			\boldmath $V_\mathrm{SW}$        & Average solar wind speed                        & 150.2$\pm$27.3  & km/s         \\
			\boldmath $B_\mathrm{0}$        & Average magnetic field                        & 0.03  &   nT     \\
			\boldmath $B$        & Magnetic field average strength                        & 0.11  &   nT     \\

			\boldmath $ n_\mathrm{p}$          & Thermal protons density                    &     (1.95$\pm$0.8) $\times10^{-3}$        & cm $^{-3}$    \\
			\boldmath $ T_\mathrm{p}$              & Thermal protons temperature                          &   (5.29$\pm$2.5)$\times10^4$      & K            \\
			\boldmath $V_\mathrm{A}$           & Alfv\'{e}n speed                      &      52.8      & km/s         \\
		    \mr
			\boldmath $r_\mathrm{cp}$          & Thermal protons Larmor radius                        &       2880     & km           \\
			\boldmath $r_\mathrm{i}$           & Thermal protons inertial radius                 &      5150        & km           \\
			\boldmath $r_\mathrm{cp~1keV}$   & 1-keV protons Larmor radius                        &       43000     & km           \\
			\mr
			\boldmath $f_\mathrm{cp~PL}$          & Thermal protons gyrofrequency  (plasma frame)              &     1.63 $\times10^{-3}$       & Hz           \\
			\boldmath $f_\mathrm{cp~SC}$          & Thermal protons Larmor frequency (SC frame)                &     2.6 $\times10^{-2}$       & Hz           \\
			\boldmath $f_\mathrm{ip~SC}$          & Thermal protons inertial frequency (SC frame)                &     1.4 $\times10^{-2}$       & Hz           \\
			\boldmath $f_\mathrm{cp~1keV~SC}$          & 1-keV protons Larmor frequency (SC frame)                &     1.7 $\times10^{-3}$       & Hz           \\
			\boldmath $f_\mathrm{e~SC}$          & 1-eddy-turnover frequency  (SC frame)             &     3.9 $\times10^{-7}$       & Hz   \\      
			\br                         
		\end{tabular}
\end{indented}
\end{table*}

The range of scales considered in this study, $f\in[10^{-8}, 10^{-2}]$ Hz, allows us to identify the large-scale, MHD, energy-injection and inertial-cascade regimes of turbulence. In principle, the transition to the kinetic regime should also be observed. In fact, the gyrofrequency of thermal protons, $f_\mathrm{cp}$, is around 1.6 mHz in the plasma reference frame, and 0.03 Hz if converted to the spacecraft-frame frequency through the Larmor radius $r_\mathrm{cp}\approx3000$ km. Ion inertial-scale structures ($r_i\approx5000$ km)  may be detected at a frequency of 0.01 Hz as well. However, as discussed in \S \ref{sec:data}, noise in the data limits the investigation to scales larger than $5\times10^{-4}$ Hz, at least at the current stage of the research.

Note that about 95\% of thermal energy of ions in the heliosheath belongs to the population of pickup-ions (PUI) \cite{malama2006,decker2008,zank2010}. Thus, PUIs are expected to have a relevant mediation effect on the turbulence, such  as that documented in \cite{smith2006a,aggarwal2016,zank2016}. The gyroradius of a 1 keV pickup proton is  about 40000 km, which may be detected in the V2 spectrum at frequencies around 1.5 mHz.

\subsection{Energy-injection regime (EI)}\label{sec:energy_inj_regime}
\begin{figure}
\centering
\includegraphics[width=\textwidth]{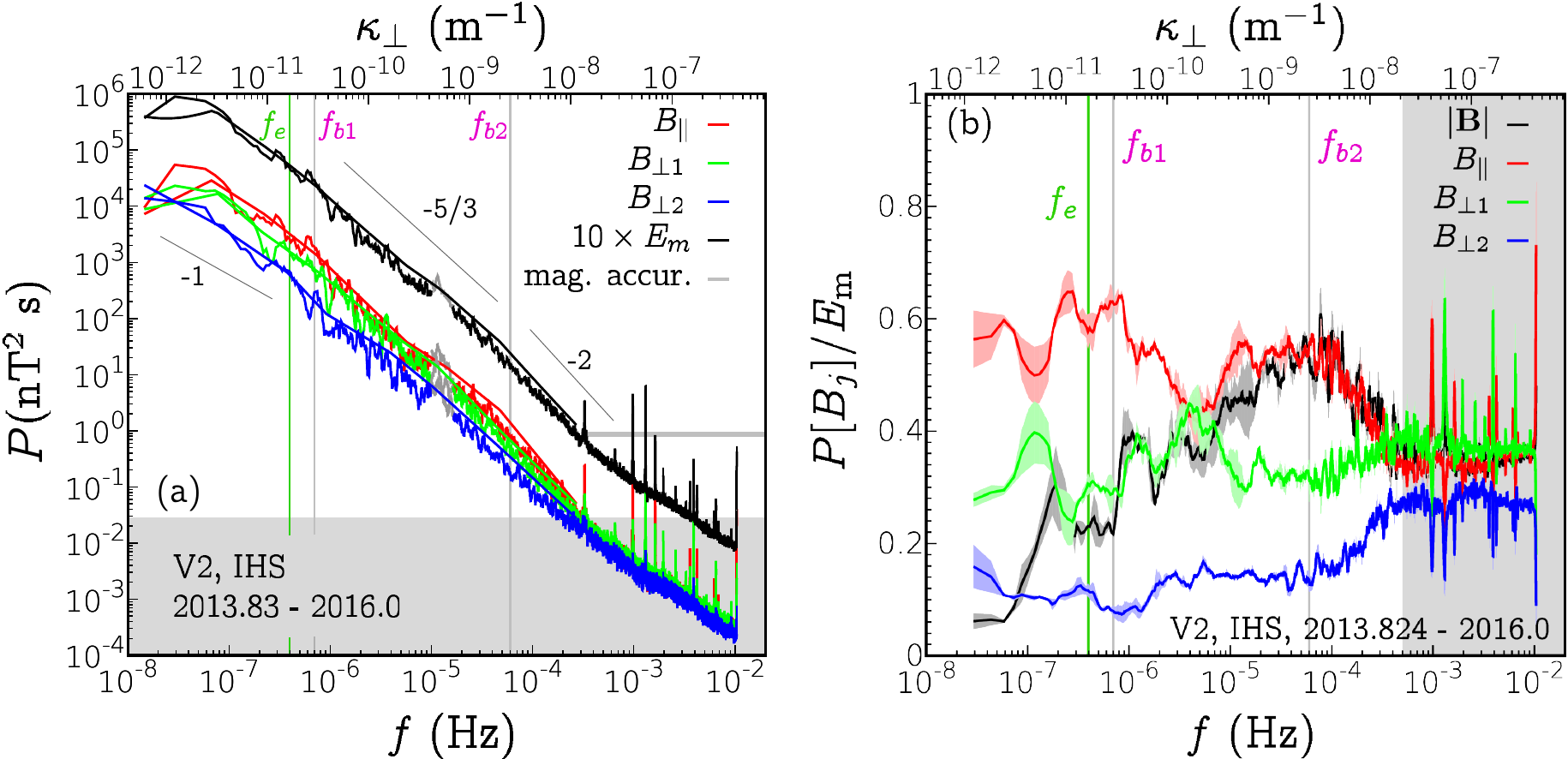}
    \caption{Power spectral density of magnetic field fluctuations, anisotropy and compressibility. (a) PSD, the three components are represented in red ($\parallel$), green ($\perp_1$), and blue ($\perp_2$). The trace is shown in black, $E_m(f)=P[B_\parallel]+P[B_{\perp1}] +P[B_{\perp2}]$.  The smooth lines show the result of the OP method, the rugged curves are build from the CI and CS methods, as described in Appendix A of \cite{fraternale2019a}. The gray band and the horizontal line represent the instrumental uncertainty threshold, modeled as a white noise with 0.03 nT amplitude per each field component. \textcolor{black}{As a reference, the three spectral slopes of -1, -5/3 and -2 are shown in the EI, IC1 and IC2 regimes, respectively}. Some sharp spikes in the PSD are instrument-related, as they are harmonics of the sampling time. The bump at 1.5$\times10^{-5}$ Hz is an artifact due to the data gaps \textcolor{black}{and it is thus shown in grey}. (b): spectral compressibility ($C(f)=P[|\mathbf{B}|]/E_m$, black curve) and spectral anisotropy for each $j$-th field component ($A_j(f)=P[B_j]/E_m$, colored curves, as in the left panel). The difference between the three spectral estimation methods is also shown. 
\label{fig:PSD}}
\end{figure}

A large-scale energy-injection regime is identified at spacecraft-frame frequencies less than $f_\mathrm{b1}\approx 7\times10^{-7}$ Hz, where a spectral break takes place, as shown in Figure \ref{fig:PSD}. This frequency corresponds to a spatial scale of about $\ell_\mathrm{b1}\approx1.3$ AU. In this regime, the power spectral density decays slowly, with a spectral index $\alpha_\mathrm{EI}\approx-1.26$. We have shown that the extension of the EI range can vary significantly in different heliosheath regions, in particular the threshold frequency is larger during unipolar periods \cite{fraternale2019a}. We also note that the EI energy decay rate observed here is faster than that observed in earlier periods. The break at $f_\mathrm{b1}$ is indeed difficult to recognize by looking at the spectrum only. However, it is clearly distinguishable from the power-law variation of the structure functions and the kurtosis shown in Figure \ref{fig:SF} ($\tau_\mathrm{b1}=1/f_\mathrm{b1}$).

In the SW upstream of the TS, this EI range with the frequency scaling of $\sim 1/f$ is found to be related to the Alfv{\'e}nicity of fluctuations and, in particular, it is more extended in fast-wind streams \cite{roberts2010}. In fact, large-scale  Alfv{\'e}nic waves in this regime did not experience yet a sufficient nonlinear interaction to produce a turbulent cascade and form a reservoir of energy for turbulence at smaller scales.

The Sun rotation provides some forcing to the system. It acts at $f_\mathrm{sun}\approx4.5\times10^{-7}$ Hz. This determines the nominal width of magnetic sectors, which is around 2 AU in the IHS. Moreover, the causality condition implies that fluctuations with spacecraft-frame frequencies below a specific threshold $f_e$ cannot be considered as ``true'' fluctuations, since they are either waves of wavelengths larger than the distance between the spacecraft and their source,  or equivalently, vortexes that did not experience yet one eddy-turnover (the typical nonlinear time scale). As a reasonable approximation for this scale we use $f_\mathrm{e}\approx {\pi V_\mathrm{SW}^2}/[{V_\mathrm{A}(r_\mathrm{SC}-r_\mathrm{source})}]$, where $r_\mathrm{SC}$ is the spacecraft location and $r_\mathrm{source}$ the location of the fluctuation's source. We consider the termination shock as a source location ($r_\mathrm{source}\approx84$ AU), which yields the value of  $f_e\approx4\times10^{-7}$ Hz reported in the figures.
If the Sun is considered as the source point, ($r_\mathrm{source}=0$),  $f_e\approx10^{-8}$~Hz. The former choice better agrees with the observed location of the large-scale spectral break. Notice also that the IHS width should be considered as the outer scale of the system \textcolor{black}{(see the sub-panel of Figure \ref{fig:SF}a)}. \textit{Voyager 1} (V1) crossed the HP at 121.5 AU, 27 AU away from the point where it crossed the TS. 

The EI regime is also identified from the structure functions of magnetic field increments as shown in Figure \ref{fig:SF}(a) for both parallel (with respect to $\mathbf{B}_0$) and perpendicular fluctuations. In fact,  the spectral break fairly well corresponds to a change in the behavior of the structure functions, \textcolor{black}{which follow flatter trends - not typical of fully-developed turbulence - for time lags $\tau>\tau_\mathrm{b1}$.} 
The EI range is not intermittent, as the kurtosis is close to Gaussian values (see the left panel of Figure \ref{fig:SF}). \textcolor{black}{At $\tau\gtrsim10^7$ s ($\ell\sim10$ AU), a peak and further flattening of $S_p$ is observed. We note that these scales are close to the outer scale of the system, but the statistic is insufficient to derive conclusions.}

It should be noted that compressibility is also small ($C\approx0.2$), as shown by the black curve in the right panel of Figure \ref{fig:PSD}. In this regard, we emphasize that the high energy of $\delta B_\parallel$ (red curve in Figure \ref{fig:PSD}) at these scales is not related to the presence of compressible modes, but rather to changes in the magnetic field polarity, clearly visible from the time history of the azimuthal angle in the second panel of Figure \ref{fig:data}. 
\textcolor{black}{In the limit of incompressible fluctuations (constant $|\mathbf{B}|$), it has been recently shown for near-Earth SW that the Kolmogorov's turbulent cascade is saturated and cannot subsist at scales, where $\mathcal{I}(\tau)$ exceeds the unity, resulting in a $\sim 1/f$ spectral power law in the EI range \cite{matteini2018}. Here, this relationship is observed in the IHS for the first time, as shown in Figure \ref{fig:SF}(a) (black curve in the sub-panel). The connection between scale-dependent compressibility and spectral regimes in the IHS was previously pointed out in \cite{fraternale2019a}, but without reference to the quantity $\mathcal{I}$.}

\FloatBarrier

\subsection{Inertial-cascade regime (IC)}\label{sec:interialcascade_regime}
\begin{figure}
\centering
\includegraphics[width=\textwidth]{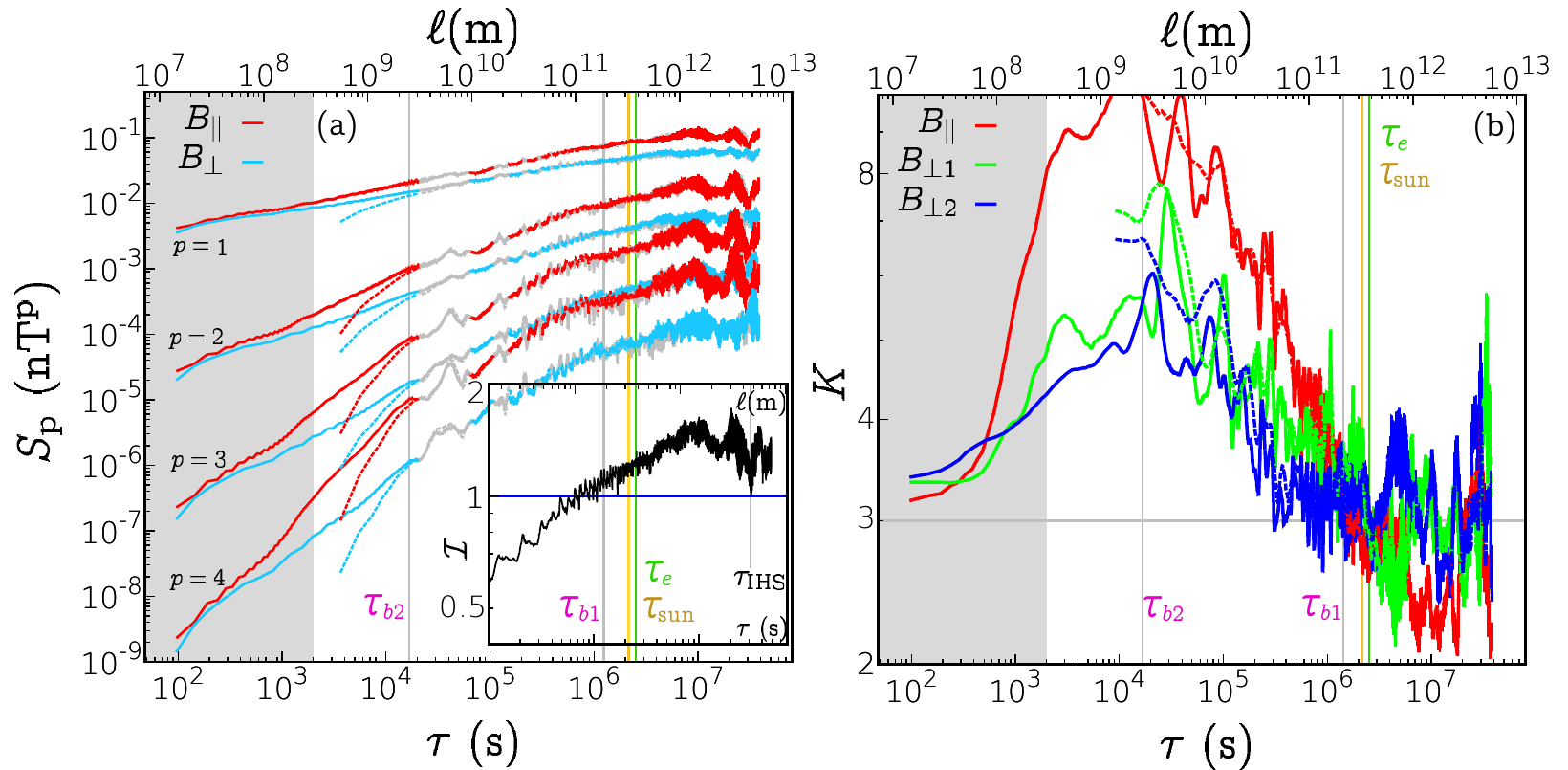}
\caption{\color{black}(a) Structure functions of magnetic field increments, $S_p$, for p=1,2,3,4. Red curves represent the structure functions of parallel fluctuations $S_{p,\parallel}$, while light-blue curves represent the average between $S_{p,\perp1}$ and  $S_{p,\perp2}$ (dashed lines represent results from 1824 s data). Oscillations are due to the data gaps in the time sequence. When the counter  $N(\tau)$ is less than 1/4 of the maximum value, the color switches to gray. The vertical lines indicate the two spectral breaks ($\tau_\mathrm{b1}$ and $\tau_\mathrm{b2}$, respectively), the 1-eddy-turnover time scale ($\tau_e$), and the Sun rotation period ($\tau_\mathrm{sun}$). The panel insert shows $\mathcal{I}(\tau)=\langle |\boldsymbol{\Delta} \mathbf{B}|/B_\tau \rangle$. Note that at the EI-to-IC spectral break $I(\tau_{b1})\approx1$, the upper limit for the turbulent cascade expected for incompressible fluctuations \cite{matteini2018}. Here, we also indicate $\ell_\mathrm{IHS}\approx V_\mathrm{SW}/\tau_\mathrm{IHS}\approx 30$ AU, that represents the IHS's width in the wind direction, under the Taylor's approximation.
(b) Scale-dependent kurtosis of magnetic increments for all the field components. It is observed that both $S_p$ and $K$ fit power laws in the inertial-cascade range between the spectral breaks ($K\sim\tau^{[-0.25,-0.14]}$).  The energy-injection range is not intermittent. \label{fig:SF} }
\end{figure}

The range of fluctuations with frequency between $f_\mathrm{b1}$ and $10^{-3}$ Hz can be referred to as the magnetohydrodynamic inertial-cascade regime of magnetic turbulence. Within IC, we highlight a second typical scale of magnetic fluctuations. In fact, it is seen that a spectral steepening takes place at  $f_\mathrm{b2}\approx6\times10^{-5}$ Hz ($\ell_\mathrm{b2}\approx2.5\times10^{6}$ km). This scale splits the IC range in two subranges which will be named  IC1 and IC2, respectively.  For $f_\mathrm{b1}<f<f_\mathrm{b2}$, we observe a defined power-law energy decay with a spectral index  $\alpha_\mathrm{IC1}\approx-1.5$ for $E_m$. This may be consistent with an anisotropic Iroshnikov--Kraichnan scaling. However, here the role of $\delta B_\parallel$ is important and they contribute mostly to the fluctuation of the magnetic field's magnitude, as shown in Figure \ref{fig:PSD}(b). The fraction of fluctuating energy due to compressible fluctuations increases from 0.2 to 0.6 in IC1, reaching the maximum at $f_\mathrm{b2}$.  The inertial cascade regime is featured by a power-law behavior of the structure functions. In particular, in the IC1 range $S_p$ show defined power laws with exponents typical of MHD turbulence (Figure \ref{fig:SF}). The exponents $\zeta_{p,\parallel},~\zeta_{p,\perp}$ are computed by linear regression in the log-log plane, excluding points with lower statistical significance due to data gaps (shown in gray in panel (a)), and are reported in Table \ref{tab:fluc_stats}. The comparison with existing theoretical models is better done computing the exponents relative to $S_3$. This was done via the \textit{extended self-similarity} principle (ESS) by which $S_p\sim S_3^{\zeta^\mathrm{ESS}}$ \cite{benzi1993}.  It is found that $S_p[S_3(\tau)]$ exhibits a significantly broader scaling range than $S_p(\tau)$, extending beyond the inertial range. This makes it possible to perform an accurate computation of the scaling exponents. The kurtosis shown in Figure \ref{fig:SF}(b) displays an increase in the IC1 range which fits the power laws $K_\parallel\sim\tau^{-0.25}$, $K_{\perp_1}\sim\tau^{-0.14}$, $K_{\perp_2}\sim\tau^{-0.19}$, where the fit of 48 s and 1824 s data gives the uncertainty of $\pm0.02$ for the exponent. The maximum intermittency is achieved around $\tau_\mathrm{b2}=1/f_\mathrm{b2}$. In the IC2 range, the intermittency reduces and Gaussian values of $K$ are found for $\tau\gtrsim2000$ s. This may be an artifact due to noise in data. However, the peak of activity at $\tau_\mathrm{b2}$ seems physical and deserves further investigation in future studies. Indeed, it is worth noticing that the power spectrum is steeper on such scales, with a spectral index $\alpha_\mathrm{IC2}\approx2.2$ for $\delta B_\parallel$ (Figure \ref{fig:PSD}). 


\begin{table*}
\lineup
\caption{Magnetic-field fluctuation properties from time-domain analysis (first six lines) and spectral analysis. \textcolor{black}{Note that in the definitions below $\delta B_j(t)=B_j(t)-B_{j,0}$, $B_0=|\mathbf{B_0}|$ is the magnitude of the average magnetic field and $B=\langle |\mathbf{B}|\rangle$ is the average magnetic field's magnitude}.\label{tab:fluc_stats} }
\begin{indented}
\centering
		\item[]\begin{tabular}{@{}llll}
			\br
		\centre{2}{Parameter} & \centre{2}{Value}                      \\
			\mr
			\boldmath $E_m$   & Average magnetic energy                        & 0.013  & nT$^2$         \\
			\boldmath $C$   & Average compressibility                        &  0.41 &          \\
			\boldmath $I_\parallel$   & Average $\parallel$ intensity $\langle \delta B_\parallel/B_0 \rangle$, \textcolor{black}{$\langle \delta B_\parallel/B \rangle$}                      &  1.97, \textcolor{black}{0.66} &          \\
			\boldmath $I_{\perp1}$   & Average $\perp_1$ intensity  $\langle \delta B_{\perp_1}/B_0 \rangle$, \textcolor{black}{$\langle \delta B_{\perp_1}/B \rangle$}                       & 1.41, \textcolor{black}{0.47}  &          \\
			\boldmath $I_{\perp2}$   & Average $\perp_2$ intensity    $\langle \delta B_{\perp_2}/B_0 \rangle$, \textcolor{black}{$\langle \delta B_{\perp_2}/B \rangle$}                      &  0.86, \textcolor{black}{0.29} &          \\
			\boldmath $I$   & Average magnitude intensity   $\langle |\mathbf{\delta B}|/B_0 \rangle$, \textcolor{black}{$\langle |\mathbf{\delta B}|/B \rangle$}                & 2.92, \textcolor{black}{0.99}  &          \\ \mr
	    	%
	    	 \boldmath $f_\mathrm{b1}$   & EI-to-IC spectral break    & $7\times10^{-7}$ & Hz\\  
	    	 \boldmath $f_\mathrm{b2}$   & IC1-to-IC2 spectral break    & $6\times10^{-5}$ & Hz\\  
        	 \boldmath $\alpha_\mathrm{EI}$   & Spectral index of $E_m$ in the EI range    & -1.26$\pm$0.07 & \\  
        	 \boldmath $\alpha_\mathrm{IC1}$   & Spectral index of $E_m$ in the IC1 range   & -1.49$\pm$0.02 & \\  
        	 \boldmath $\alpha_\mathrm{IC2}$   & Spectral index of $E_m$ in the IC2 range   & -1.87$\pm$0.07 & \\  \mr
        	 \boldmath $\zeta_{1,\parallel} (\zeta_{1,\parallel}^\mathrm{ESS}$)   & 1st SF-exponent of $\delta B_\parallel$     & 0.35 (0.45) & \\  
        	 \boldmath $\zeta_{2,\parallel}~(\zeta_{2,\parallel}^\mathrm{ESS})$   & 2nd SF-exponent of $\delta B_\parallel$     & 0.62 (0.78) & \\  
        	 \boldmath $\zeta_{3,\parallel}(\zeta_{3,\parallel}^\mathrm{ESS})$   & 3rd SF-exponent of $\delta B_\parallel$     & 0.80 (1) & \\  
        	 \boldmath $\zeta_{4,\parallel}~(\zeta_{4,\parallel}^\mathrm{ESS})$   & 4th SF-exponent of $\delta B_\parallel$     & 0.95 (1.16) & \\  \mr
             \boldmath $\zeta_{1,\perp}~(\zeta_{1,\perp}^\mathrm{ESS})$   & 1st SF-exponent of $\delta B_\perp$     & 0.27 (0.39) & \\  
             \boldmath $\zeta_{2,\perp}~(\zeta_{2,\perp}^\mathrm{ESS})$   & 2nd SF-exponent of $\delta B_\perp$     & 0.50 (0.73) & \\  
             \boldmath $\zeta_{3,\perp}~(\zeta_{3,\perp}^\mathrm{ESS})$   & 3rd SF-exponent of $\delta B_\perp$     & 0.71 (1) & \\  
             \boldmath $\zeta_{4,\perp}~(\zeta_{4,\perp}^\mathrm{ESS})$   & 4th SF-exponent of $\delta B_\perp$     & 0.89 (1.22) & \\  
			\br                         
		\end{tabular}
\end{indented}
\end{table*}

\color{black}
\section{Conclusions}
This study reports on the \textit{Voyager 2} observations of magnetic turbulence in the inner heliosheath from 2013.824 to 2016.0, when V2 was at 106.5$\pm$3.4 AU from the Sun. It is believed that during this period the spacecraft was inside the sector region of the inner heliosheath. The present paper shows follow-up results of our recent work \cite{fraternale2019a}, where the spectral properties of magnetic field fluctuations have been shown for a collection of several unipolar and sector IHS regions, and for LISM intervals, for a spectral bandwidth over six decades.

We identify two scales that may be characteristic to the turbulence in this region. The first scale is located at $f_\mathrm{b1}\approx7\times10^{-7}$ Hz (spacecraft-frame frequency), which corresponds to structures of size $\ell_\mathrm{b1}\approx{1.3}$ AU in the solar wind direction - under the Taylor's approximation. This scale seems discriminating the energy-injection range of magnetic field fluctuations from a second regime which can be interpreted as the inertial-cascade range of turbulence. The first regime consists of incompressible and non-intermittent fluctuations, following a power law for the reduced power spectra with spectral index $\alpha_\mathrm{EI}\approx-1.25$. At these scales, the magnitude of magnetic field increments is between one and two times the average magnetic field's strength.

The inertial-cascade regime displays a spectral index about $\alpha_\mathrm{IC1}\approx-1.5$ and power-law growing intermittency. Here, we observe a second scale located at $f_{b2}\approx6\times10^{-5}$ Hz ($\ell_\mathrm{b2}\approx0.017$ AU), where a faster cascade is observed until $5\times10^{-4}$ Hz, especially in the B-parallel fluctuations ($\alpha_\mathrm{IC2}\approx-2$). Concurrently, here the maximum compressibility and intermittency are observed. Higher frequencies, up to 0.01 Hz, should include the transition from the magnetohydrodynamic regime to the kinetic regime. However, this range could be affected by noise in the data and was not considered in the present discussion. 
The influence of energetic particle populations as well as the potential presence of turbulent magnetic reconnection should be considered in future research in order to shed light into the nature of these observations.
\color{black}

\FloatBarrier

\ack F.F. acknowledges support from the postdoctoral grant “FOIFLUT” 37/17/F/AR-B. N.P. was supported, in part, by NASA grants NNX14AJ53G, NNX16AG83G, 80NSSC18K1649NS, and 80NSSC19K0260, and NSF PRAC award OAC-1811176. J.D.R. was supported under NASA contract 959203 from JPL to MIT. Computational resources for spectral and statistical analysis were provided by HPC@POLITO (\url{http://www.hpc.polito.it}).

\section*{References}

\begin{thebibliography}{10}
\expandafter\ifx\csname url\endcsname\relax
  \def\url#1{{\tt #1}}\fi
\expandafter\ifx\csname urlprefix\endcsname\relax\def\urlprefix{URL }\fi
\providecommand{\eprint}[2][]{\url{#2}}

\bibitem{fraternale2019a}
Fraternale F, Pogorelov N~V, Richardson J~D and Tordella D 2019 {\em Astrophys.
  J.} {\bf 872} 40

\bibitem{mccomas2006}
McComas D~J and Schwadron N~A 2006 {\em Geophys. Res. Lett.\/} {\bf 33} L04102

\bibitem{fisk2009}
Fisk L~A and Gloeckler G 2009 {\em Adv. Space Res.\/} {\bf 43} 1471--1478

\bibitem{heerikhuisen2010}
Heerikhuisen J, Pogorelov N~V, Zank G~P, Crew G~B, Frisch P~C, Funsten H~O,
  Janzen P~H, McComas D~J, Reisenfeld D~B and Schwadron N~A 2010 {\em
  Astrophys. J. Lett.\/} {\bf 708} L126--L130

\bibitem{pogorelov2016}
Pogorelov N~V, Bedford M~C, Kryukov I~A and Zank G~P 2016 {\em J. Phys. Conf.
  Series\/} {\bf 767} 012020

\bibitem{zhao2017}
Zhao L~L, Adhikari L, Zank G~P, Hu Q and Feng X~S 2017 {\em Astrophys. J.\/}
  {\bf 849} 88

\bibitem{hill2014}
Hill M~E, Decker R~B, Brown L~E, Drake J~F, Hamilton D~C, Krimigis S~M and
  Opher M 2014 {\em Astrophys. J.\/} {\bf 781} 94

\bibitem{richardson2015}
Richardson J~D and Decker R~B 2015 {\em J. Phys. Conf. Ser.\/} {\bf 577} 012021

\bibitem{pogorelov2017b}
Pogorelov N~V, Heerikhuisen J, Roytershteyn V, Burlaga L~F, Gurnett D~A and
  Kurth W~S 2017 {\em Astrophys. J.\/} {\bf 845} 9

\bibitem{richardson2013}
Richardson J~D and Burlaga L~F 2013 {\em Space Sci. Rev.\/} {\bf 176} 217--235

\bibitem{zank2015}
Zank G~P 2015 {\em Annu. Rev. Astron. Astrophys.
\/} {\bf 53} 449

\bibitem{pogorelov2017a}
Pogorelov N~V, Fichtner H, Czechowski A, Lazarian A, Lembege B, Roux J~A~I,
  Potgieter M~S, Scherer K, Stone E~C, Strauss R~D, Wiengarten T, Wurz P, Zank
  G~P and Zhang M 2017 {\em Space Sci. Rev.\/} {\bf 212} 193--248

\bibitem{burlaga2006}
Burlaga L~F, Ness N~F and Acuna M~H 2006 {\em Astrophys. J.\/} {\bf 642} 584

\bibitem{burlaga2006b}
Burlaga L~F, Ness N~F and Acuna M~H 2006 {\em J. Geophys. Res.-Space Physics\/}
  {\bf 111} A09112

\bibitem{burlaga2009}
Burlaga L~F and Ness N~F 2009 {\em Astrophys. J.\/} {\bf 703} 311

\bibitem{burlaga2015}
Burlaga L~F, Florinski V and Ness N~F 2015 {\em Astrophys. J. Lett.\/} {\bf
  804} L31

\bibitem{burlaga2018}
Burlaga L~F, Florinski V and Ness N~F 2018 {\em Astrophys. J.\/} {\bf 854} 20

\bibitem{drake2010}
Drake J~F, Opher M, Swisdak M and Chamoun J~N 2010 {\em Astrophys. J.\/} {\bf
  709} 963--974

\bibitem{drake2017}
Drake J~F, Swisdak M, Opher M and Richardson J~D 2017 {\em Astrophys. J.\/}
  {\bf 837} 159

\bibitem{pogorelov2013b}
Pogorelov N~V, Borovikov S~N, Bedford M~C, Heerikhuisen J, Kim T~K, Kryukov I~A
  and Zank G~P 2013 Modeling solar wind flow with the multi-scale fluid-kinetic
  simulation suite {\em Numerical modeling of space plasma flows
  astronum-2012\/} vol 474 ed Pogorelov N~V, Audit E and Zank G~P pp 165--171
  7th Annual International Conference on Numerical Modeling of Space Plasma
  Flows, Big Isl, HI, JUN 25-29, 2012

\bibitem{pogorelov2012}
Pogorelov N~V, Borovikov S~N, Zank G~P, Burlaga L~F, Decker R~A and Stone E~C
  2012 {\em Astrophys. J. Lett.\/} {\bf 750} L4

\bibitem{pogorelov2013a}
Pogorelov N~V, Suess S~T, Borovikov S~N, Ebert R~W, McComas D~J and Zank G~P
  2013 {\em Astrophys. J.\/} {\bf 772} 2

\bibitem{pogorelov2015}
Pogorelov N~V, Borovikov S~N, Heerikhuisen J and Zhang M 2015 {\em Astrophys.
  J. Lett.\/} {\bf 812} L6

\bibitem{kim2017}
Kim T~K, Pogorelov N~V and Burlaga L~F 2017 {\em Astrophys. J. Lett.\/} {\bf
  843} L32

\bibitem{burlaga2018b}
Burlaga L~F, Ness N~F and Richardson J~D 2018 {\em Astrophys. J.\/} {\bf 861} 9

\bibitem{burlaga2017}
Burlaga L~F, Ness N~F and Richardson J~D 2017 {\em Asptrophys. J.\/} {\bf 841} 47

\bibitem{richardson2016b}
Richardson J~D and the Voyager~Team 2016 Voyager observations in the outer heliosphere and interstellar medium AIP Conf. Proc. vol 1720, {\em SOLAR WIND 14: Proceedings of the Fourteenth International Solar Wind Conference}, ed. L. Wang et al. (Melville, NY: AIP), 080001

{\em Solar Wind 14\/} ({\em AIP Conf.
  Proc.\/} vol 1720)

\bibitem{behannon1977}
Behannon K~W, Acuna M~H, Burlaga L~F, Lepping R~P, Ness N~F and Neubauer F~M
  1977 {\em Space Sci. Rev.\/} {\bf 21} 235--257

\bibitem{fraternale2016}
Fraternale F, Gallana L, Iovieno M, Opher M, Richardson J~D and Tordella D 2016
  {\em Physica Scripta\/} {\bf 91} 394--401

\bibitem{iovieno2016}
Iovieno M, Gallana L, Fraternale F, Richardson J~D, Opher M and Tordella D 2016
  {\em European Journal of Mechanics B/Fluids\/} {\bf 55} 394--401

\bibitem{gallana2016}
Gallana L, Fraternale F, Iovieno M, Fosson S~M, Magli E, Opher M, Richardson
  J~D and Tordella D 2016 {\em J. Geophys. Res. - Space Physics\/} {\bf 121}
  3905--3919 

\bibitem{sorriso2019}
Sorriso-Valvo L, Catapano F, Retinò A, Le~Contel O, Perrone D, Roberts O~W,
  Coburn J~T, Panebianco V, Valentini F, Perri S, Greco A, Malara F, Carbone F,
  Veltri P, Pezzi O, Fraternale F, Di~Mare F, Marino F, Giles B, Moore T~E, T
  R~C, Torbert R~B, Burch J~L and Khotyaintsev Y~V 2019 {\em Phys. Rev.
  Lett.\/} {\bf 122} 035102

\bibitem{fraternale2017phd}
Fraternale F 2017 {\em Internal waves in fluid flows. Possible coexistence with
  turbulence\/} Ph.D. thesis Politecnico di Torino

\bibitem{berdichevsky2009}
Berdichevsky D~B 2009 Voyager Mission, Detailed processing of weak
magnetic fields; II - Update on the cleaning of Voyager magnetic field
density B with MAGCALs (Washington, DC: NASA)

\bibitem{roberts1987a}
Roberts D~A, Klein L~W, Goldstein M~L and Matthaeus W~H 1987 {\em J. Geophys.
  Res.\/} {\bf 92} 11021--11040

\bibitem{taylor1938}
Taylor G~I 1938 {\em Proc. R. Soc. Lond. A\/} {\bf 164} 476--490  

\bibitem{zhou2004}
Zhou Y, Matthaeus W~H and Dmitruk P 2004 {\em Rev. Mod. Phys.\/} {\bf 76}
  1015--1035

\bibitem{malama2006}
Malama Y~G, Izmodenov V~V and Chalov S~V 2006 {\em Astron. Astrophys.\/} {\bf
  445} 693--701

\bibitem{decker2008}
Decker R~B, Krimigis S~M, Roelof E~C, Hill M~E, Armstrong T~P, Gloeckler G,
  Hamilton D~C and Lanzerotti L~J 2008 {\em Nature\/} {\bf 454} 67--70

\bibitem{zank2010}
Zank G~P, Heerikhuisen J, Pogorelov N~V, Burrows R and McComas D 2010 {\em
  Astrophys. J.\/} {\bf 708} 1092--1106

\bibitem{smith2006a}
Smith C~W, Hamilton K, Vasquez B~J and Leamon R~J 2006 {\em Astroph. J.
  Lett.\/} {\bf 645} L85--L88

\bibitem{aggarwal2016}
Aggarwal P, Taylor D~K, Smith C~W, Joyce C~J, Fisher M~K, Isenberg P~A, Vasquez B~J, Schwadron N~A, Cannon B~E and Richardson J~D 2016 {\em Astrophy. J.\/}{\bf 822} 94

\bibitem{zank2016}
Zank P~G, 2016 {\em Geosci. Lett.\/}{\bf 3} 22
  {\bf 822} 94

\bibitem{matteini2018}
Matteini L, Stansby D, Horbury T~S, Chen C~H~K 2018 {\em Astrophys. J. Lett.\/} {\bf 869} L32
  

\bibitem{roberts2010}
Roberts D~A 2010 {\em J. Geophys. Res.\/} {\bf 115} A12101

\bibitem{benzi1993}
Benzi R, Ciliberto S, Tripiccione R, Baudet C, Massaioli F and Succi S 1993
  {\em Phys. Rev. E.\/} {\bf 48} R29--R32

















\end{thebibliography}

\end{document}